\documentclass[a4paper,9pt]{article}

\usepackage{graphicx}

\begin{document}

%\vspace*{2.3cm}

%\textbf{Velocity fields of distant galaxies with FORS2
\textbf{Velocity fields of distant galaxies with FORS2 at the VLT
}

\bigskip

Bodo L. Ziegler$^1$,\\
Elif Kutdemir$^2$,\\
Cristiano Da Rocha$^1$,\\
Asmus B\"ohm$^3$,\\
Wolfgang Kapferer$^3$,\\
Harald Kuntschner$^1$,\\
Reynier F. Peletier$^2$,\\
Sabine Schindler$^3$\\
Miguel Verdugo$^4$

\bigskip

$^1$ESO\\
$^2$Kapteyn Institute, Groningen, Netherlands\\
$^3$University Innsbruck, Austria\\
$^4$University G\"ottingen, Germany

\bigskip

\textbf{We describe a method to efficiently obtain two-dimensional velocity 
fields of 
distant, faint and small, emission-line galaxies with FORS2 at the VLT.
They are examined for kinematic substructure to identify possible interaction
processes.
Numerical simulations of tidal interactions and ram-pressure effects
reveal distinct signatures observable with our method.
We detect a significant fraction of galaxies with irregular velocity fields
both in the field and cluster environments.
}

\bigskip

Galaxy formation and evolution are still principal research topics in modern
astrophysics although having been investigated since the recognition of objects
outside the Milky Way after the Great Debate between Shapley and Curtis.
The main questions include:
When and how did galaxies form?
Why are there different morphological types?
What is the role of environment?
Models range from single epoch to continuous creation scenarios.
The current theoretical paradigm of a cold dark matter dominated universe
predicts a hierarchical bottom-up structure evolution from small entities
towards bigger systems through merging.
During such a process spiral galaxies can be transformed into ellipticals,
however, the disc component can be regained through subsequent new gas 
accretion and star formation.
Many irregular and peculiar galaxies are observed in clusters of galaxies,
in which they may experience interactions in addition to merging and accretion.

Much progress was achieved in recent years by deep photometric surveys
(like GOODS)
finding galaxies and quasars at high redshift ($z\sim6$) corresponding to an 
epoch just a billion years after the Big Bang and revealing a dichotomy between
blue star forming and red passive galaxies at later times with an evolutionary
transition between the two groups.
Spectroscopic surveys
(like CFHRS)
additionally produced evidence that there is a sharp decline in the overall
star formation activity in the cosmos within the last eight billion years.
However, all these results are based on measurements of the luminosity as
produced by the stars, while the total mass of a galaxy is mainly composed of 
dark matter.
Because the triggering and process of star formation contains 
complicated physics, we need to make many assumptions and simplifications 
in our modeling, and, in addition, we do not clearly understand the 
scaling between the baryonic and dark matter.
Therefore, it would be more favourable if we could observationally determine 
the total mass of galaxies and compare its evolution directly to the predicted
structure assembly.

Such measurements of the dynamical mass even of faint and small, very distant
galaxies were achieved recently by a number of research groups using the
largest telescopes.
Through spectroscopy they derive the internal kinematics (motions) of the 
stellar system that are subject to the whole gravitational potential.
In the optical wavelengths regime, we can conduct this out to redshifts of 
unity.
For spiral galaxies, we can determine the rotation curve from gas
emission lines
(like [O II] at 372.7nm restframe wavelength).
If the galaxy is undisturbed, the rotation curve rises in the inner
part and turns over into a flat plateau dominated by its dark halo and we can
calculate the maximum rotation velocity $V_{\rm max}$.
In such a case the assumption of virialization holds, so that we can use the
Tully-Fisher relation TFR 
(Tully \& Fisher 1977)
as a powerful diagnostic tool that scales the
baryonic matter (parameterised, e.g., by stellar luminosity) to the dark 
matter (as given by $V_{\rm max}$).
Our group, for example, has found that the TFR of 130 distant 
(z=0.5 on average)
field galaxies has a shallower slope than the local one if restframe B-band
luminosities are considered, so that the brightening of a slowly rotating 
spiral is much larger than the one of a fastly rotating galaxy
(B\"ohm \& Ziegler 2007).
This result can be interpreted principally in two ways:
either the majority of distant galaxies are much more scattered around the
TFR and we do not see enough regular spirals at the faint end
or there is a mass-dependent evolution with low-mass objects exhibiting
on average a stronger effect over the last five billion years.
The latter scenario is supported by chemical evolution modeling of our galaxies
that revealed less efficient but extended star formation for slow rotators
(Ferreras et al. 2004)
and is consistent with other observational evidence for a mass-dependent
evolution (often called ``downsizing'').

For a correct application of the TF analysis it is indispensable that a galaxy
is undisturbed so that the assumption of virial equilibrium is fulfilled and,
therefore, the rotation curve shows a very regular shape.
%However, what can we learn from all the peculiar cases?
If this is not the case, can we then still learn something?
In the CDM structure formation theory, galaxies evolve by merging and 
accretion, so that we can look for kinematic signatures of such events and 
whether there is an increase in the abundance of such types with redshift.
To investigate a possible environmental dependence, we need to disentangle
such collisions from numerous other interaction phenomena galaxies can
experience in groups and clusters.
The gas content of spirals reacts to the ram-pressure exerted by the 
intracluster medium, a hot plasma permeating the whole structure.
Part of the gas can also be stripped off by tidal forces 
(sometimes called ``strangulation'').
Other processes like harassment can even remove stars from galaxy discs, so
that the overall morphology gets altered.
However, (irregular) rotation curves from slit spectroscopy do not contain 
unique information to reveal unambigously a specific interaction process.
In addition, peculiar shapes can sometimes be caused artificially
by observational and instrumental effects when long-slits are used
(Ziegler et al. 2003, J\"ager et al. 2004).
A more favourable observation would, therefore, be that of a two-dimensional
velocity field.

\bigskip 

\textbf{FORS2 spectroscopy to obtain velocity fields
}

\bigskip 

Many large telescopes today offer 3D-spectroscopy and most future facilities
(like E-ELT)
will provide such instrument modes.
At the VLT, for example, VIMOS and FLAMES offer IFU spectroscopy in the
optical regime.
For our specific purpose, however, we conceived a different method to obtain
velocity fields that has many advantages for our science case:
matched galaxy sizes, good spatial resolution, long wavelength range and
high efficieny through a large number of simultanous targets and economic 
exposure times
(Ziegler et al. 2007, Kutdemir et al. 2008).

For our procedure, we utilise the MXU mode of FORS2 that allows to cut slits
individually and at any desired orientation into a mask by a laser.
We pick up a particular target and start with a slit placing it along the 
photometric major axis as measured on spatially highly resolved HST/ACS images
(Figure 1).
The slit covers the full size of the galaxy in its mid-plane with width 1''
and extends much beyond it to allow an accurate sky subtraction 
(common slit lengths are 15-25'').
In the same manner, 20-30 more slits are placed on other targets across the
full field of view of FORS2 (6.8' x 6.8');
occasionally even two or three objects fall into the same slit.
Observations of such a mask will yield in the end rotation curves like for our
previous projects with one spatial axis and one velocity axis by measuring
centre positions of an emission line row-by-row along the spatial profile of
the two-dimensional galaxy spectrum
(B\"ohm et al. 2004).
In order to construct a velocity field with two spatial axes we observe all
targets twice again with different slit positions using two more masks.
This time slits are placed 1'' offset along the minor axis of
the galaxy to either side of the first position, so that all three slit
positions together correspond to a rectangular grid
(Figure 1).
That way, the full extent of a galaxy is covered, which is particularly
important for a TF galaxy, where we need to measure the flat part of the
rotation curve, which is usually reached at about three disc scale lengths.
At the redshifts of our targets (0.1 - 1.0), this corresponds to angular
sizes of about 3'' to 6''.
The spatial resolution along the x-axis of our grid is given by the pixel
size of the FORS2 CCD chips of 0.25''.
For the y-axis, we are restricted by the slit width of 1.0'', which is a good
compromise between minor losses of light not falling into the slit due to 
seeing, adequate spectral resolution (R=1000),
and spatial sampling in the y-direction.

As spectral element we use a high-throughput holographic VPH grism (600RI)
that results in a long wavelength range (330nm) projected onto the CCD
with two advantages:
for each galaxy several gas emission lines are visible
and many absorption lines of the stellar continuum can be combined
to derive the stellar rotation curve in addition to the gaseous one.
Depending on the redshift of the object, emission lines from the blue [O II]
line (at 372.7nm) to the red Halpha line (at 656.3nm) are visible.
While up to seven different lines can be observed in case of some field 
galaxies, most cluster members have four
(Figure 2).
This allows us to derive rotation curves and velocity fields in several
independent ways and to check for consistency but also to
investigate possible dependencies on the physical state of the gas clouds
that emit the respective lines
(all galaxies have both recombination and forbidden lines).
From line ratios, we can then construct metallicity maps and assess a possible
contribution from AGN activity in addition to star formation.

Despite the need for the observation of three masks to construct a 
velocity field, our method is still highly efficient.
The VPH grism with its medium spectral resolution together with the excellent
optics of FORS2 makes it possible that with 2.5 hours total integration time,
sufficient signal is reached in an emission line to measure accurately line
centres in each spectral (CCD) row.
This holds also for the outer regions of the galaxies, where the surface
brightness drops to low values 
(from typically 22$V$mag/arcsec$^2$ in the inner part to about 
26$V$mag/arcsec$^2$).
Therefore, we can obtain the necessary data of a high number of objects 
(20-30) within a total integration time of only 7.5 hours.
For comparison, FLAMES observations with 15 IFUs of similar targets need 
integration times of 8-13 hours.

The standard reduction of the spectroscopic images produces a wavelength 
calibrated two-di\-men\-sional spectrum for each slit position of the mask
separately.
Thanks to the high stability of FORS, exposures of the same mask taken in
different nights can be combined to a single deep spectrum.
Like for the determination of rotation curves, the wavelength of the centre
position of a given emission line is measured along the spatial profile.
Differences to a common systemic centre can then be translated into velocity
space applying Doppler's rule.
The construction of the velocity field VF needs a very careful combination
of the measurements from the three slit positions.
In order to know the exact position of the masks relative to each other, we
also included a few stars with narrow slits perpendicular to each other.
These stellar spectra are also used to determine the seeing during the 
exposure.
In addition, only for a third of the galaxies per mask the central slit 
position was taken while the other two thirds had the off-centre positions
alleviating both mask acquisition during observations and mask combination
after data reduction.
To setup the VF, each position-velocity data point is then calculated within
a shared coordinate system, whose origin was determined from the intensity 
maximum
of the spatial profile around the emission line in the central slit and its
respective wavelength.
An example is shown in Figure 3.

\bigskip 

\textbf{Kinemetric analysis of velocity fields
}

\bigskip 

Since our VFs cover a large fraction of a galaxy's extent with good spatial
resolution, we can analyse them quantitatively to some detail.
We use kinemetry
(Krajnovic et al. 2006)
that was originally developed for nearby
galaxies observed with the SAURON 3D-spectrograph, whose VFs have much higher 
signal and resolve much smaller physical scales than in the case of our
distant targets.
In polar coordinates the velocity profile of a flat rotating disc can be 
described by a cosine function of the azimuthal angle.
Best fitting ellipses, along which the velocity profiles are extracted, can be 
determined as a function of radius.
Deviations from these fits can be quantified by a harmonic Fourier expansion, 
whose coefficients can be interpreted in terms of physical parameters.
While the first order reflects the bulk motion (the rotation), the third and 
fifth order, for example, describe the correction to simple rotation and
indicate separate kinematic components.
For the reconstruction of the intrinsic velocity map
free parameters are the flattening and position angle of the ellipses, 
while their centres are fixed to a common origin.
An example of a reconstructed velocity map with best-fitting ellipses
overplotted is given in Figure 4.
When, instead, the inclination and position angle are fixed to their average
global value, a simple rotation field along the kinematic major axis can be 
modeled. 
Its residuals with respect to the observed VF may indicate noncircular 
velocity components like streaming motions along spiral arms or along a bar.  
Additional components such as a decoupled core can be recognized in such a 
residual map, too, but also by twists in the position angle and flattening as
well as an increase in the $k_5$ Fourier coefficient. 
A rotation curve is extracted from the observed VF along the kinematic major 
axis to accurately determine $V_{\rm max}$, which may be different from the
one derived from the curve along the central slit aligned with the photometric
axis.

%\bigskip 
\vspace{2cm}

\textbf{Simulations of interactions: structure \& kinematics
}

\bigskip 

One of the main goals of our project is to identify signatures of possible
interaction events.
Galaxies may be transformed from one type into another by merging or
accretion but also by other effects like ram-pressure stripping or
harassment in the environment of a cluster or group.
Open questions are still, which processes are efficient under what conditions
and whether there is a dominant mechanism responsible for the abundance of
elliptical galaxies in local rich clusters.
In order to study systematically the distortions and irregularities in
rotation curves and velocity fields caused by interaction phenomena, we perform
N-body/SPH simulations and extract both structural and kinematic information
from computer output in the same manner as from observational data.
The numerical calculations are made for the three components dark matter halo,
stellar body and collisional gas clouds of a galaxy
and are based on the Gadget2 code
(Springel 2005)
that incorporates hydrodynamic physics
(and has explicit prescriptions for star formation and feedback).
So far, we have modeled minor and major mergers, tidal interactions caused
by fly-bys, and ram-pressure stripping
(Kronberger et al. 2006, 2007, 2008).
There are large variations in the degree of distortions both in stellar 
structure and gas kinematics in case of the first three processes with
main dependencies on the mass ratio of the galaxies, the geometry of the
interaction in real space and the projection onto the sky plane (the viewing
angle of the observations).
In the case of a major merger, however, there is always a clear signal
in the higher-order coefficients of kinemetry's Fourier decomposition
(Figure 5).
%(Figure 5a).
Ram-pressure by the intracluster medium, on the other hand, mainly affects the 
gas disc pushing away the outer parts creating distortions at the edge of VFs,
while regular rotation can be maintained in the inner regions
(Figure 6).
%(Figure 5b).
In certain configurations, the gas disc can also be displaced from the centre
of the stellar disc.
During the stripping event, gas can also be compressed so that enhanced
star formation is triggered, leading to changes in the stellar populations, too
(Kapferer et al. 2009).

In addition, we utilise the simulations in order to systematically 
investigate how velocity fields and characteristic parameters are influenced
by the instrumental and observational setup (like spectral resolution,
seeing) and for which cases artificial distortions can be induced
(Kapferer et al. 2006).
A major impact is caused by the decreasing spatial resolution of the spectra
when galaxies are observed at higher and higher redshifts.
Minor irregularities can be completely smeared out due to seeing and
resolution elements being too coarse.

Currently, we are in the process of analyzing each observed galaxy 
individually.
In case of a regular VF, we perform a TF analysis deriving the rotation curve
along the kinematic major axis.
For irregular galaxies we examine both the VF and the stellar structure 
(as revealed on our HST/ACS images)
and compare them to a suite of simulated events taking into account the 
specific galaxy parameters in order to pin down the specific interaction 
mechanism that caused the distortions.

\bigskip 

%\textbf{Global analysis of irregularities}
\textbf{Velocity fields of distant galaxies
}

\bigskip 

The spectroscopic observations for the presented method were performed in
Periods 74 \& 75 in four different cluster fields
(MS1008.1-1224 z=0.30, 
MS2137.3-2353 z=0.31, 
Cl0412-65 z=0.51, 
MS0451.6-0305 z=0.54).
The target galaxies were not only cluster members but also field galaxies in
the background and foreground ($0.1<z<0.9$),
so that we can investigate different environments.
All cluster fields were imaged with ACS onboard HST
enabling an accurate assessment of morphological structure of the galaxies
as well.
Our method presented here yielded velocity fields for 49 objects with
good signal-to-noise appropriate for our kinemetric analysis.
Out of these, there are 16 VFs suitable for our investigation
of possible interaction processes in rich clusters.

In addition to our detailed analysis with comparison to the simulations,
we also investigate average properties to assess the abundance of galaxies 
that have irregular kinematics irrespective of any assumption of a 
particular interaction process.
To that purpose, we quantify deviations from a simple smooth rotation field with
three different indicators measured for each galaxy in the same way.
As pure gas kinematic tracers we use 
1) $\sigma_{PA}$: the standard deviation of the kinematic position angles
of the best-fitting ellipses found by kinemetry across a galaxy
and 2) $k_{3,5}/k_1$: an average value of higher-order Fourier 
coefficients normalised by the rotation velocity.
The third parameter compares the global velocity field determined by spectral
lines emitted by the (warm) gas content of a galaxy to its morphological
structure seen in the continuum light of the stars:
3) $\Delta\phi$: the mean difference between photometric and kinematic
position angles across a galaxy.
To find an appropriate limit of the value range, below which a galaxy can 
still be classified as undistorted, we measured the three parameters first
for a local sample (taken from Daigle et al. 2006) of 18 galaxies that have
high-resolution VFs
(Kutdemir et al. 2008).

For our distant sample, we find that the
fraction of galaxies classified to be irregular according to the three
indicators is not unique.
For the pure kinematic tracers we derive a much lower percentage
(about 10\%\ and 30\%) 
than for the third parameter (about 70\%)
(Figure 7 and Kutdemir et al. 2009).
%(Figure 6 and Kutdemir et al. 2009).
The parameters trace different signatures of external processes but are also
sensitive to intrinsic properties.
For example, the presence of a bar misaligned to the disc's major axis can
also cause a large offset between the gas field and the stellar component.
In addition, peculiarities in a VF are more affected by a coarse
spatial binning and could be ``smeared out''.
Modeling resolution effects we found that the irregularity fractions we 
measure are lower limits only.

Furthermore, our simulations show that only strong interactions like
major mergers induce large values of the kinematic tracers
indicating big distortions with high significance.
So, most of our observed objects with smaller values of the irregularity
parameters probably undergo more subtle events.
This is presumably also the reason for the surprisingly similar abundance of 
peculiar galaxies in the field and cluster environment no matter what 
indicator is chosen.
%Clusterspecific interactions are either less important or less frequent,
Since we detect more irregular field galaxies at intermediate redshifts than
in the local sample, we probably witness the ongoing growth of their discs via 
accretion and minor mergers as predicted in CDM models.
To decisively clarify the ongoing processes both for field and cluster 
galaxies, we will in the near future examine all available pieces of 
information for each galaxy
(gas VFs, stellar rotation curves, morphologies, stellar populations)
and compare them to our simulations.

\bigskip 

Acknowledgements: This work was financially supported by
VolkswagenStiftung (I/76 520), DFG (ZI 663/6), 
DLR (50OR0602, 50OR0404, 50OR0301) and Kapteyn institute.
Based on observations ESO PID 74.B--0592 \& 75.B-0187 
as well as HST PID 10635.

\bigskip 

B\"ohm et al. 2004 A\&A 420, 97\\
B\"ohm \& Ziegler 2007 ApJ 668, 846\\
Ferreras et al. 2004 MNRAS 355, 64\\
Daigle et al. 2006 MNRAS 367, 469\\
J\"ager et al. 2004 A\&A 422, 941\\
Kapferer et al. 2006 A\&A 446, 847\\
Kapferer et al. 2009 A\&A 499, 87\\
Krajnovi\'{c} et al. 2006 MNRAS 366, 787\\
Kronberger et al. 2006 A\&A 458, 69\\
Kronberger et al. 2007 A\&A 473, 761\\
Kronberger et al. 2008 A\&A 483, 783\\
Kutdemir et al. 2008 A\&A 488, 117\\
Kutdemir et al. 2009 A\&A in prep.\\
%Kutdemir et al. 2009 A\&A submitted\\
Springel 2005 MNRAS 364, 1105\\
Tully \& Fisher 1977 A\&A 54, 661\\
Ziegler et al. 2003 ApJL 598, 87\\
Ziegler et al. 2007 IAU Symp. 235, 258

\bigskip 

\begin{figure}[h]
\centerline{
\resizebox{12cm}{!}{\includegraphics{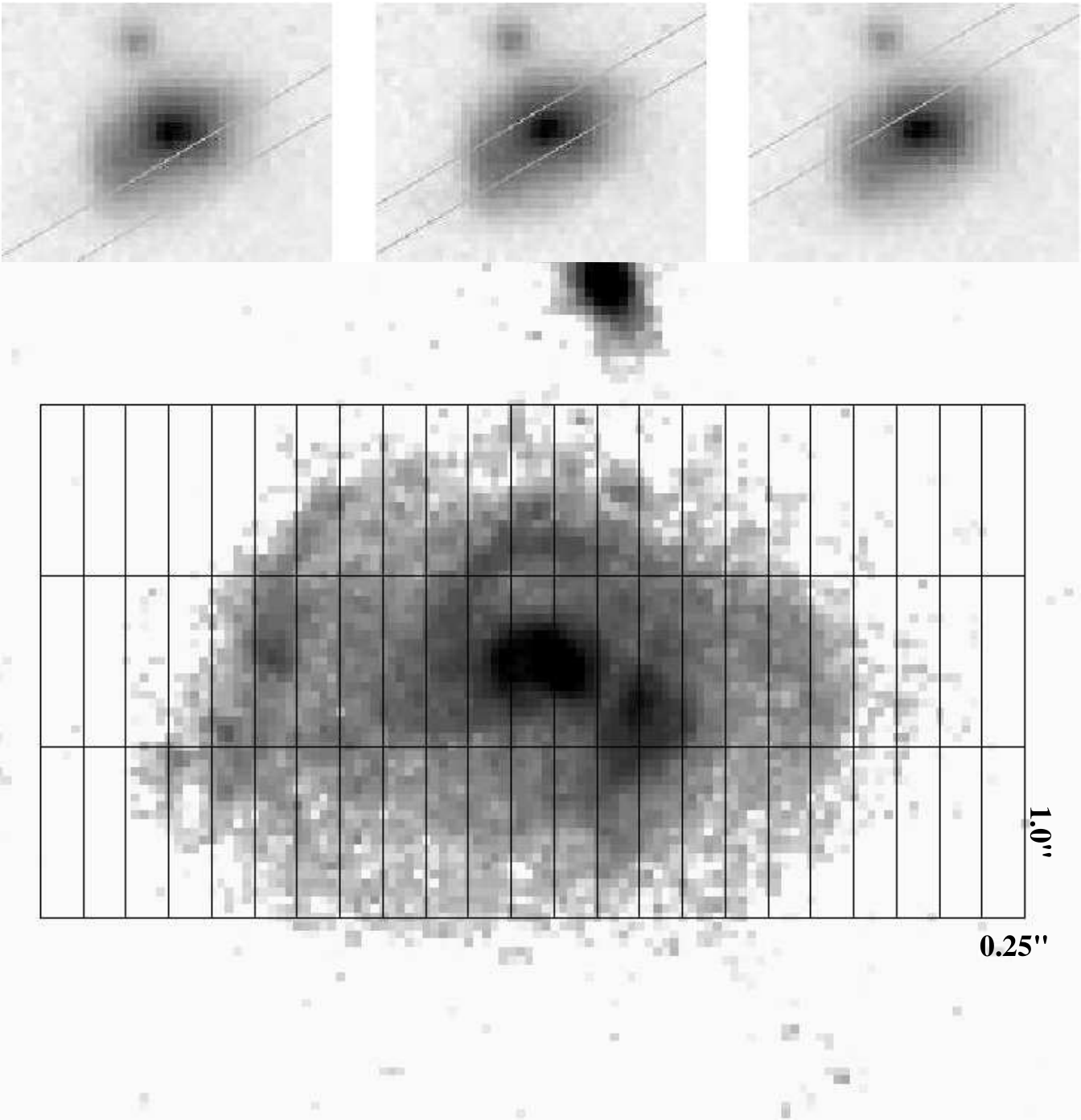}}
}
\caption{Our method to obtain two dimensional velocity fields with FORS2
is based on the observation of three different slit positions per target.
The combination of the measurements yield independent data points for a
rectangular grid covering the whole galaxy
(from Kutdemir at al. 2008).
}
\end{figure}

\begin{figure}[h]
\centerline{
\resizebox{12cm}{!}{\includegraphics{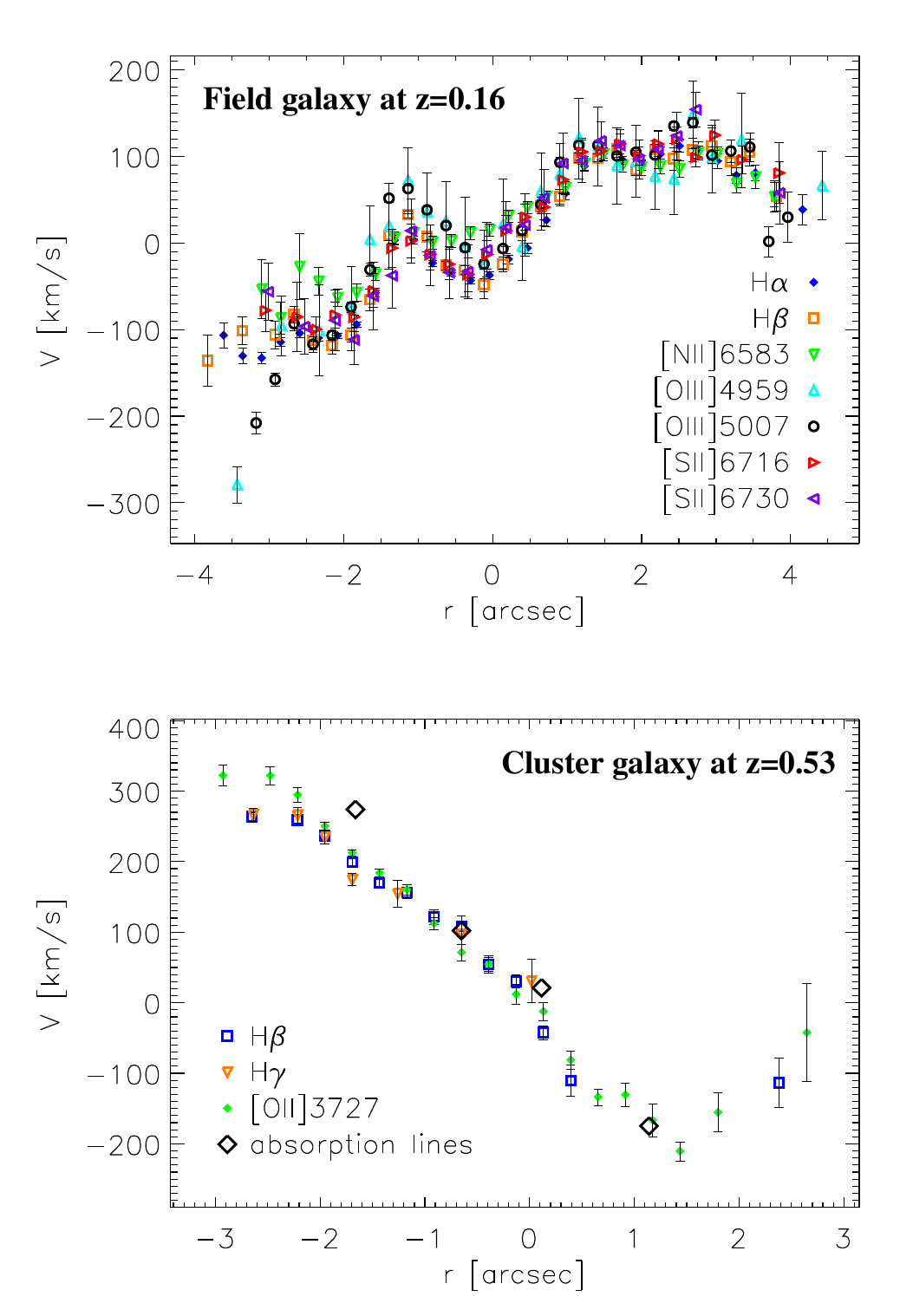}}
}
\caption{Two examples of rotation curves derived from the central slit.
For some field galaxies up to seven different emission lines can be used
independently to study the gas kinematics.
For brighter galaxies, even stellar rotation curves can be measured by
combining many absorption lines
(from Kutdemir at al. 2008).
}
\end{figure}

\begin{figure}[h]
\centerline{
\resizebox{12cm}{!}{\includegraphics{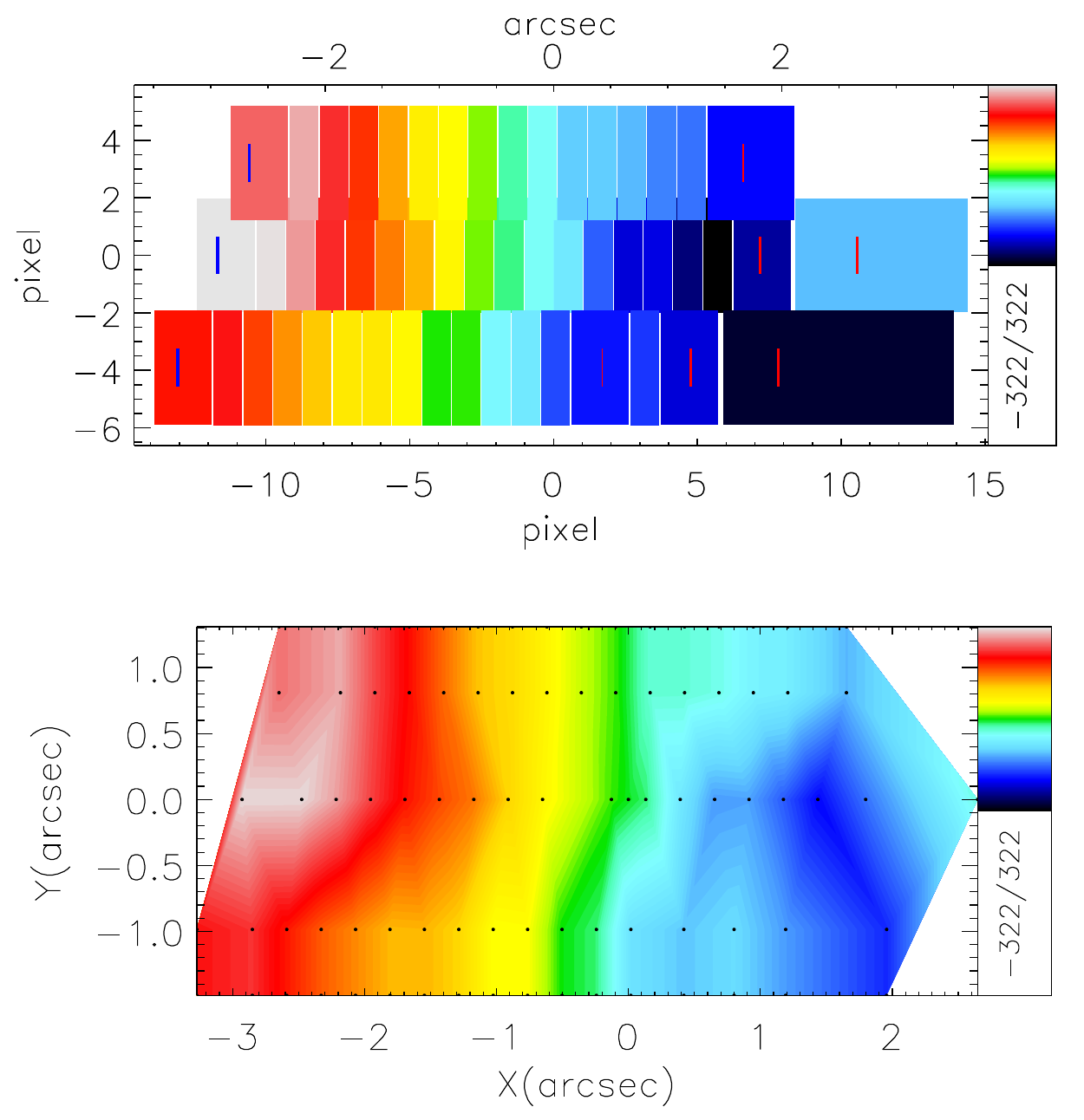}}
}
\caption{Example of an observed velocity field of a cluster spiral at z=0.5
displayed as binned independent data points (top panel)
and linearily interpolated for visualisation purposes (bottom panel)
(from Kutdemir at al. 2008).
}
\end{figure}

\begin{figure}[h]
\centerline{
\resizebox{14cm}{!}{\includegraphics{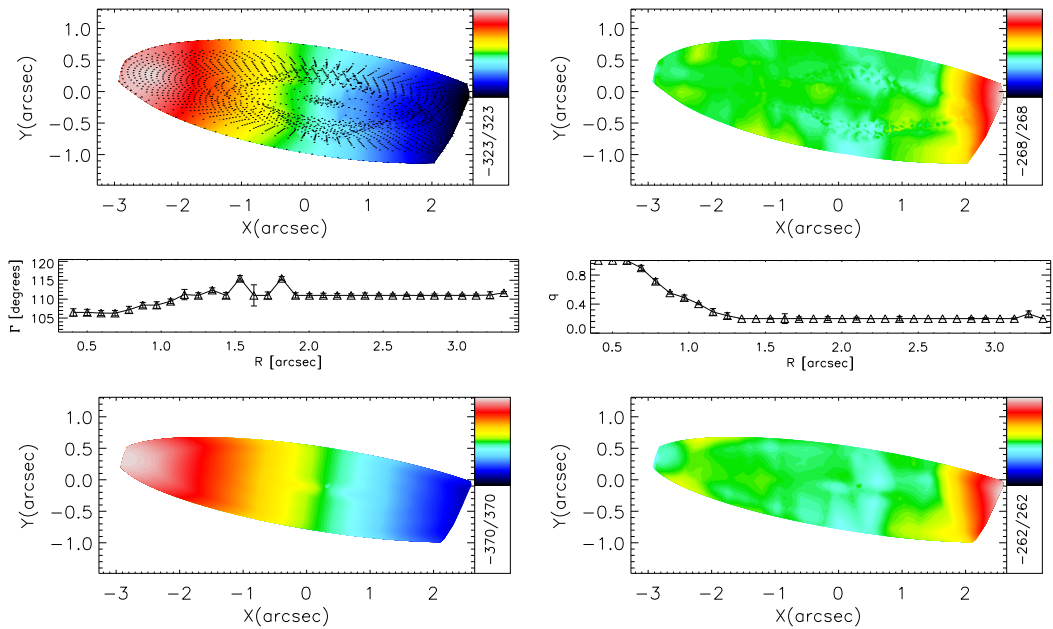}}
}
\caption{Velocity field of galaxy of Figure 3 reconstructed by kinemetry
with best-fitting ellipses overplotted (top left) and its residual map (top 
right) obtained by subtracting the model from the observed field.
Displayed in bottom panels is the model
(left) and its residual (right) of the circular velocity
component (rotation map) constructed for the average value of the kinematic
position angle $\Gamma$ and flattening $q$.
Radial profiles of the latter parameters are shown in middle panels as function
of distance to kinematic centre
(from Kutdemir at al. 2008).
}
\end{figure}

\begin{figure}[h]
\centerline{
\resizebox{12cm}{!}{\includegraphics{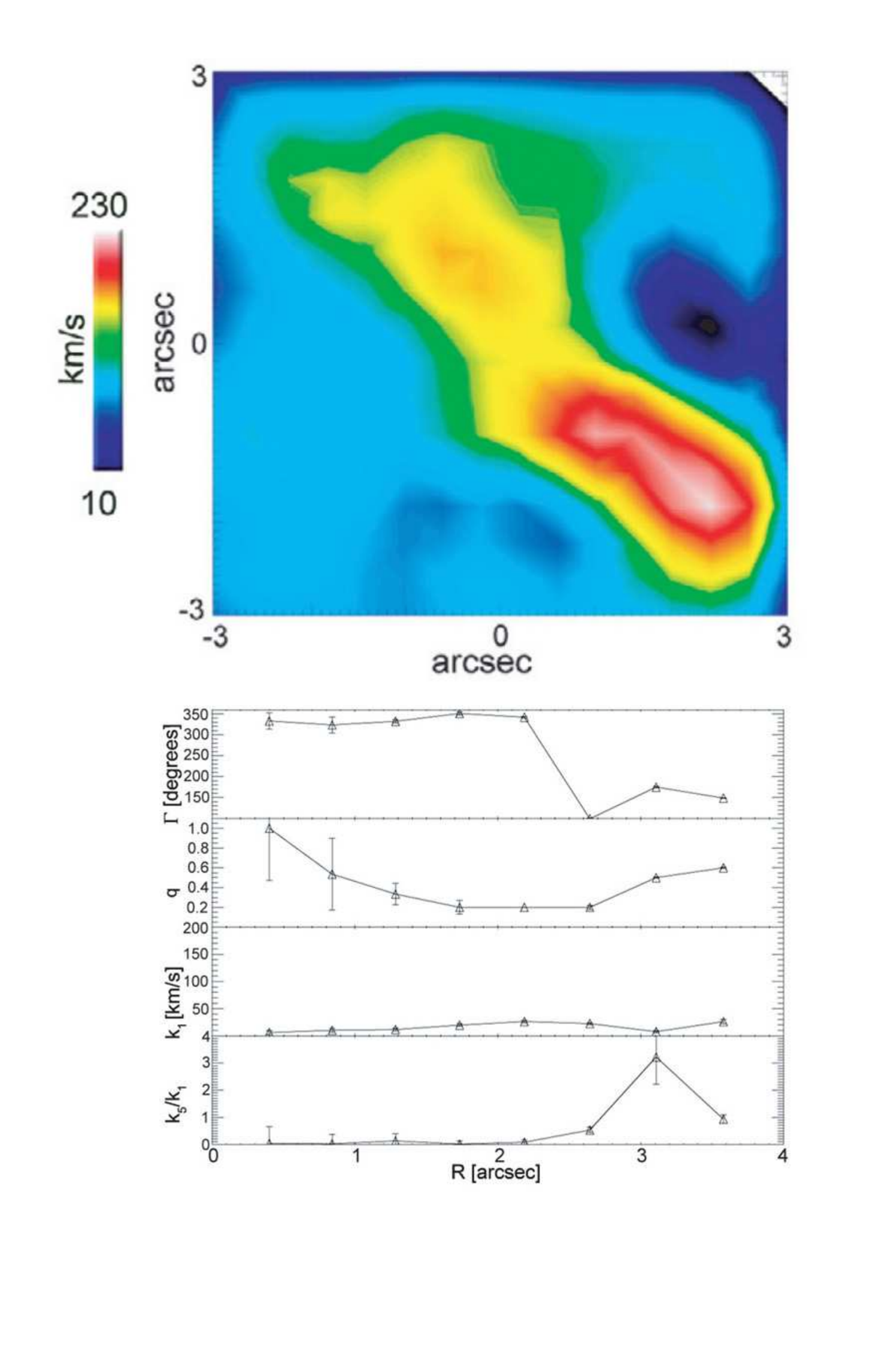}}
}
%\caption{Figure 5a:
\caption{
Velocity field of a simulated major merger 
with spatial resolution corresponding to observations at z=0.1 
and radial profiles of some parameters determined by kinemetry
(kinematic position angle, flattening, first Fourier coefficient indicating
the bulk motion and normalised fifth Fourier coefficient indicating separate
kinematic components).
The case shown refers to a small galaxy that has penetrated a bigger one,
200Myr after the event
(from Kronberger at al. 2007).
}
\end{figure}

\begin{figure}[h]
\centerline{
\resizebox{15cm}{!}{\includegraphics{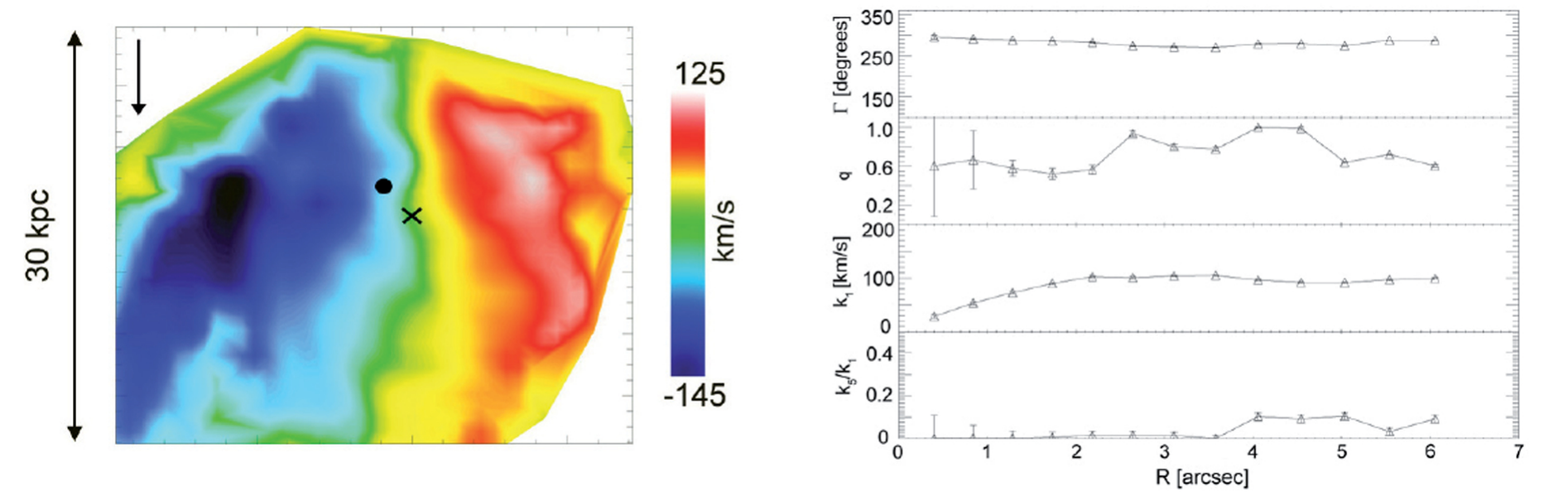}}
}
%\caption{Figure 5b:
\caption{
Velocity field of a simulated ram-pressure stripped galaxy 
with spatial resolution corresponding to observations at z=0.1 
and radial profiles of some parameters determined by kinemetry
(kinematic position angle, flattening, first Fourier coefficient indicating
the bulk motion and normalised fifth Fourier coefficient indicating separate
kinematic components).
The case shown refers to edge-on ram-pressure acting already for 400Myr
with black arrow indicating ICM wind direction and cross and circle the centre
of kinematic and stellar disc, respectively
(from Kronberger at al. 2008).
}
\end{figure}

\clearpage

\begin{figure}[h]
\centerline{
\resizebox{14cm}{!}{\includegraphics{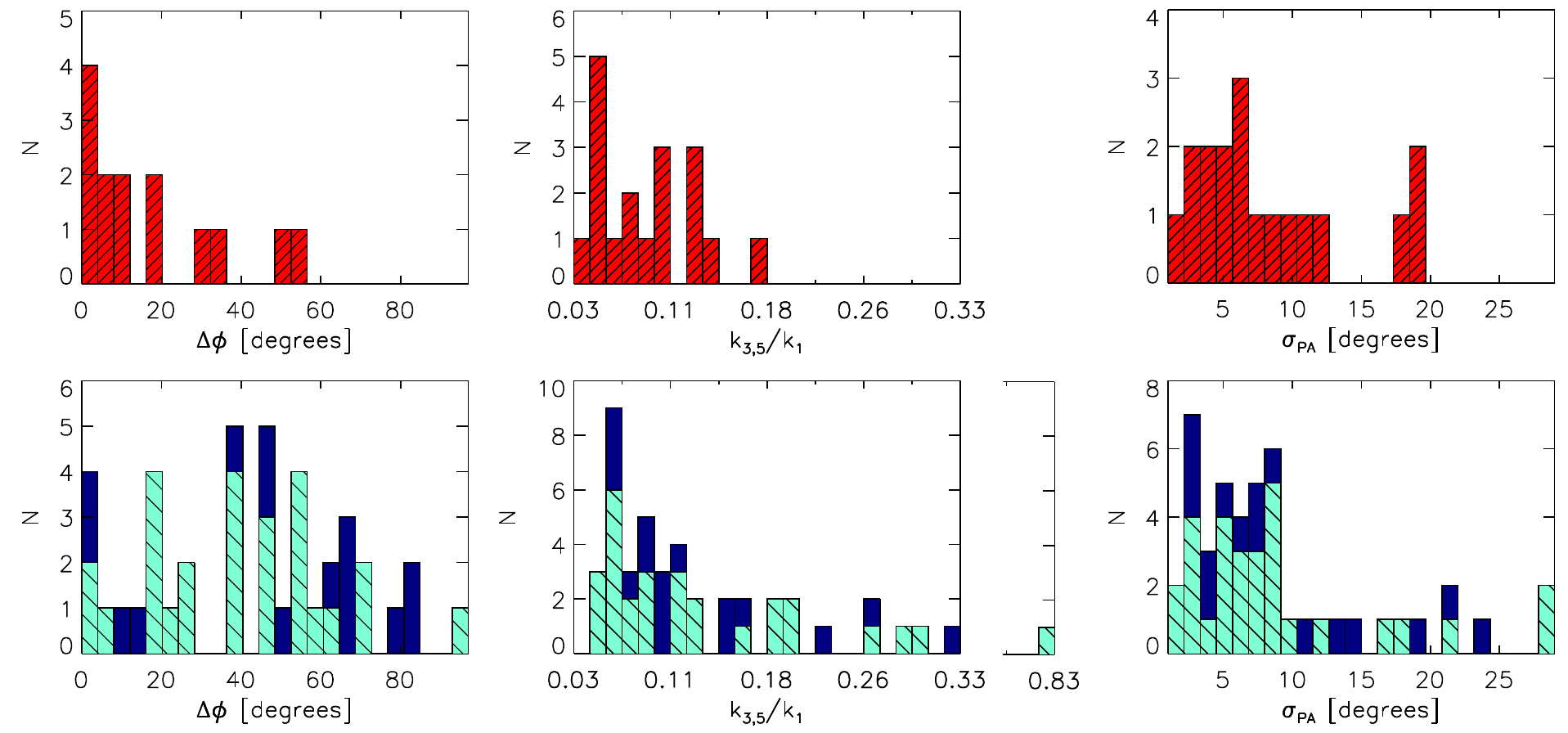}}
}
%\caption{Figure 6:
\caption{
Abundance of galaxies distributed according to our three irregularity 
parameters.
Top panels: local sample, bottom panels: distant field (green hashed) and
cluster galaxies (blue filled histograms).
Galaxies are classified irregular if
$\Delta\phi>25$, $k_{3,5}/ k_{1}>0.15$ or $\sigma_{PA}>20$
(from Kutdemir at al. 2009).
}
\end{figure}

\end{document}